\documentclass[aps,pra,superscriptaddress,amsfonts,amssymb,amsmath,eqsecnum,nofootinbib,twocolumn,floatfix]{revtex4-2}
  \usepackage{color,graphicx}
  \usepackage[utf8]{inputenc}
 \usepackage[T2A]{fontenc}
  \usepackage[english]{babel}
  \usepackage{ulem}
  \definecolor{darkblue}{rgb}{0,0,0.7}
 \definecolor{darkred}{rgb}{0.7,0,0}
 \definecolor{darkgreen}{rgb}{0,0.4,0}
 \usepackage[unicode, colorlinks, citecolor=darkblue, linkcolor=darkred, urlcolor=blue]{hyperref}

 \allowdisplaybreaks
\begin{document}

\author{Aleksandr V. Karpenko}

\affiliation{Faculty of Physics, M.V. Lomonosov Moscow State University, Leninskie Gory, Moscow 119991, Russia}

\author{Andrey B. Matsko}

\affiliation{Jet Propulsion Laboratory, California Institute of Technology, 4800 Oak Grove Drive, Pasadena, California 91109-8099, USA}

\author{Sergey P. Vyatchanin}
\affiliation{Faculty of Physics, M.V. Lomonosov Moscow State University, Leninskie Gory, Moscow 119991, Russia}
\affiliation{Faculty of Physics, Branch of M.V. Lomonosov Moscow State University in Baku,\\ 1 Universitet street, Baku, AZ1144, Azerbaijan}

\date{\today}
	
\title{Optical Entanglement Facilitated by a Hot Mechanical Oscillator}

\begin{abstract}
Optomechanical generation of entangled optical beams is usually hindered by thermal noise. We present a theoretical study of low frequency entanglement generation between two nondegenerate optical harmonics emitted from a cavity optomechanical system operating in the resolved-sideband regime. The system comprises three nearly equidistant optical modes in a high-finesse Fabry–Perot cavity, with the central mode coherently driven by a continuous wave classical signal. This configuration enables radiation-pressure interactions that generate strong quantum correlations between the two undriven sideband modes. Remarkably, these correlations persist even at large numbers of thermal phonons and a broad range of system parameters if the optical pump is strong enough. Entanglement is verified using a full quantum Langevin model that accounts for thermal noise and optical losses. Our findings demonstrate the feasibility of robust entanglement under ambient conditions, opening new avenues for hybrid quantum technologies based on mechanical interfaces and continuous-variable quantum information processing. \copyright 2026 All Rights Reserved.
\end{abstract}

\maketitle

\section{Introduction}

Entanglement is one of the most profound and nonclassical features of quantum mechanics, and it has played a central role in discussions of the foundations of physics since the early work of Einstein, Podolsky, and Rosen (EPR) \cite{einstein1935can}. Their seminal argument, pointing to the apparent incompleteness of quantum theory, motivated decades of research into the nature of quantum correlations. Schrodinger subsequently coined the term ``entanglement'' to emphasize its fundamental significance \cite{schrodinger1935discussion}. The development of Bell's inequalities \cite{bell1964einstein} and their experimental violation \cite{aspect1982experimental,aspect1982bell,haensel1990violation} established entanglement as a genuine physical phenomenon rather than a paradoxical artifact. Today, entanglement is understood not only as a signature of nonlocality, but also as a resource enabling quantum communication, quantum computation, and precision metrology \cite{horodecki2009quantum}.

In parallel with these conceptual advances, continuous-variable (CV) entanglement has emerged as a practical framework for implementing quantum information protocols using quadrature-amplitude correlations of bosonic modes of light and matter \cite{braunstein2005quantum}. The Duan–Simon inseparability criterion \cite{duan2000inseparability,simon2000peres} and the Reid EPR criterion for inferred variances \cite{reid1989demonstration} provided experimentally testable conditions to verify CV entanglement. Since then, entangled optical beams have been generated using parametric down-conversion \cite{ou1992realization}, four-wave mixing \cite{josse2004generation}, and atomic ensembles \cite{julsgaard2001experimental}.

More recently, cavity optomechanics has emerged as a powerful platform for studying CV entanglement at the interface of light and matter \cite{aspelmeyer2014cavity}. In such systems, the quantized radiation pressure force couples optical fields to the motion of a mechanical element, enabling a wide range of nonlinear quantum phenomena. Optomechanical interactions can, in principle, mediate entanglement between mechanical and optical modes \cite{vitali2007optomechanical}, between multiple mechanical oscillators \cite{mancini2002entangling,riedinger2018remote,ockeloen2018stabilized}, and between distinct optical channels \cite{pirandola2003entanglement, genes2008robust, 2011BarzanjehPRA, 2013BarzanjehPRA}. This versatility makes optomechanics a natural candidate for hybrid quantum networks that combine disparate degrees of freedom as well as for CV quantum information protocols operating at telecommunication or microwave frequencies.

Experimental progress has already demonstrated squeezed light from optomechanical cavities \cite{safavi2013squeezed,purdy2013strong,purdy2017optomechanical} and even entanglement between distant mechanical oscillators mediated by light \cite{riedinger2018remote,ockeloen2018stabilized}. However, realizing robust and stationary entanglement between multiple optical harmonics mediated by a mechanical degree of freedom remains challenging, particularly under ambient conditions. The central difficulty is the omnipresent thermal noise that populates mechanical resonators. Thermal fluctuations act as an effective decoherence channel, washing out quantum correlations and suppressing entanglement \cite{vitali2007optomechanical,aspelmeyer2014cavity,miki2026amplification}. For this reason, most experimental demonstrations of optomechanical entanglement have required cryogenic environments, transient protocols, high-quality mechanical oscillators, or active reservoir engineering \cite{2013ClerkPRA,vanner2011pulsed,hofer2011quantum,genes2008robust,woolley2014two}. 

Remarkably, steady-state entanglement between two independently pumped optical cavities in a thermalized three-mode optomechanical configuration has recently been demonstrated \cite{2015ClerkPRA,2019BarzanjehNature}. The system achieves this despite the coupling of the optical cavities to a single mechanical resonator mode populated by a large number of thermal phonons, and without requiring auxiliary cryogenic cooling or feedback control. Here, we show that a simpler and more general four-mode optomechanical (or electro-optical) system can be described within the same theoretical framework as the three-mode configuration. We further analyze possible experimental schemes for entanglement detection and compare their efficiencies.

We study theoretically a three-mode Fabry–Perot cavity configuration operating in the resolved-sideband regime. The cavity has one of the mirrors movable with the free motion having properties of a linear harmonic oscillator. The system includes three nearly equidistant optical resonances (with frequencies $\omega_+$, $\omega_0$, and $\omega_-$), with $\omega_0$ by frequency of the central mode ($\omega_0$) coherently driven by an external laser. Frequencies $\omega_\pm$ are separated from $\omega_0$ by frequency $\omega_m$ of mechanical oscillator: $\omega_\pm= \omega_0\pm \omega_m$. Radiation pressure couples the driven mode to a mechanical oscillator, and through this interaction generates strong quantum correlations between the two undriven sideband modes. 

We show that the quantum correlations persist even in the presence of strong thermal noise. Using a full quantum Langevin treatment that incorporates both optical losses and thermal decoherence, we demonstrate that the output sideband modes can become entangled under realistic parameters. Our results highlight the feasibility of generating robust CV entanglement of optical beams in ambient conditions, thereby paving the way for practical opto-mechanical interfaces for quantum communication, sensing, and information processing.
  
Experimental verification of this entanglement requires separation of the output light corresponding to the modes $\omega_\pm$. However, direct spectral discrimination is challenging, since the frequency difference $\omega_+ - \omega_-$ is negligibly small on the optical scale. To circumvent this limitation, we propose the use of synodyne detection \cite{Buchmann2016}. In this approach, the combined output from all three modes, $\omega_0$ and $\omega_\pm$, is directed to a balanced homodyne detector, while the local oscillator is prepared with two carrier frequencies, $\omega_+$ and $\omega_-$. By adjusting the relative amplitudes and phases of these two frequency components of the local oscillator, it is possible to select arbitrary weighted combinations of quadratures. This capability provides the necessary measurement flexibility to reveal the presence of entanglement between the optical sidebands.  Our results indicate that, under realistic experimental conditions, the proposed detection scheme enables direct observation of optomechanical entanglement between the sideband modes $\omega_\pm$.

The proposed optomechanical configuration has several notable advantages. First, it represents a generic architecture that does not require multiple optical cavities, thereby reducing the complexity of the system. Second, entanglement generation remains possible even when the optical modes have unequal coupling strengths and decay rates, demonstrating robustness against parameter mismatch. Third, the scheme is compatible with a broad range of optomechanical geometries beyond the specific configuration considered here. Finally, an equivalent system with the same mathematical structure can be realized in a resonant electro-optical platform, such as a dual-sideband resonant electro-optic modulator, suggesting a broader applicability of the proposed entanglement-generation mechanism. Fifth, the generated entanglement is promising for improvement of detection of classical signals and potentially can improve resonant gravity wave detectors sensitivity.

\section{Results}

\subsection{Physical system and Hamiltonian}
The unitary part of the system under study is governed by the canonical interaction Hamiltonian 
$H_\text{int} \sim (E_0 + E_+ + E_-)^2 x$, where $E_0$, $E_-$, and $E_+$ are the electric fields of the modes $0$, $-$, and $+$ on the surface of the moving mirror of a Fabry–Perot resonator. The modes can belong to different mode families to ensure that there are no other optical harmonics interacting with the mechanical oscillator. Within the rotating wave approximation, the Hamiltonian of the system can be written as
\begin{eqnarray}
  && H   = H_0 + H_\text{int},\\
  && H_0 =\hslash \omega_+\hat c_+^\dag \hat c_+  + \hslash \omega_0 \hat c_0^\dag \hat c_0 
    + \hslash \omega_-\hat c_-^\dag \hat c_-
         +\hslash \omega_m \hat d^\dag \hat d, \nonumber \\ \nonumber
    && H_\text{int}  = -i\hslash
      \left( \left[\eta_-\hat c_0^\dag \hat c_- + \eta_+\hat c_+^\dag \hat c_0\right] \hat d -adj.\right)\nonumber
\end{eqnarray}
\begin{figure}
 \includegraphics[width=0.48\textwidth]{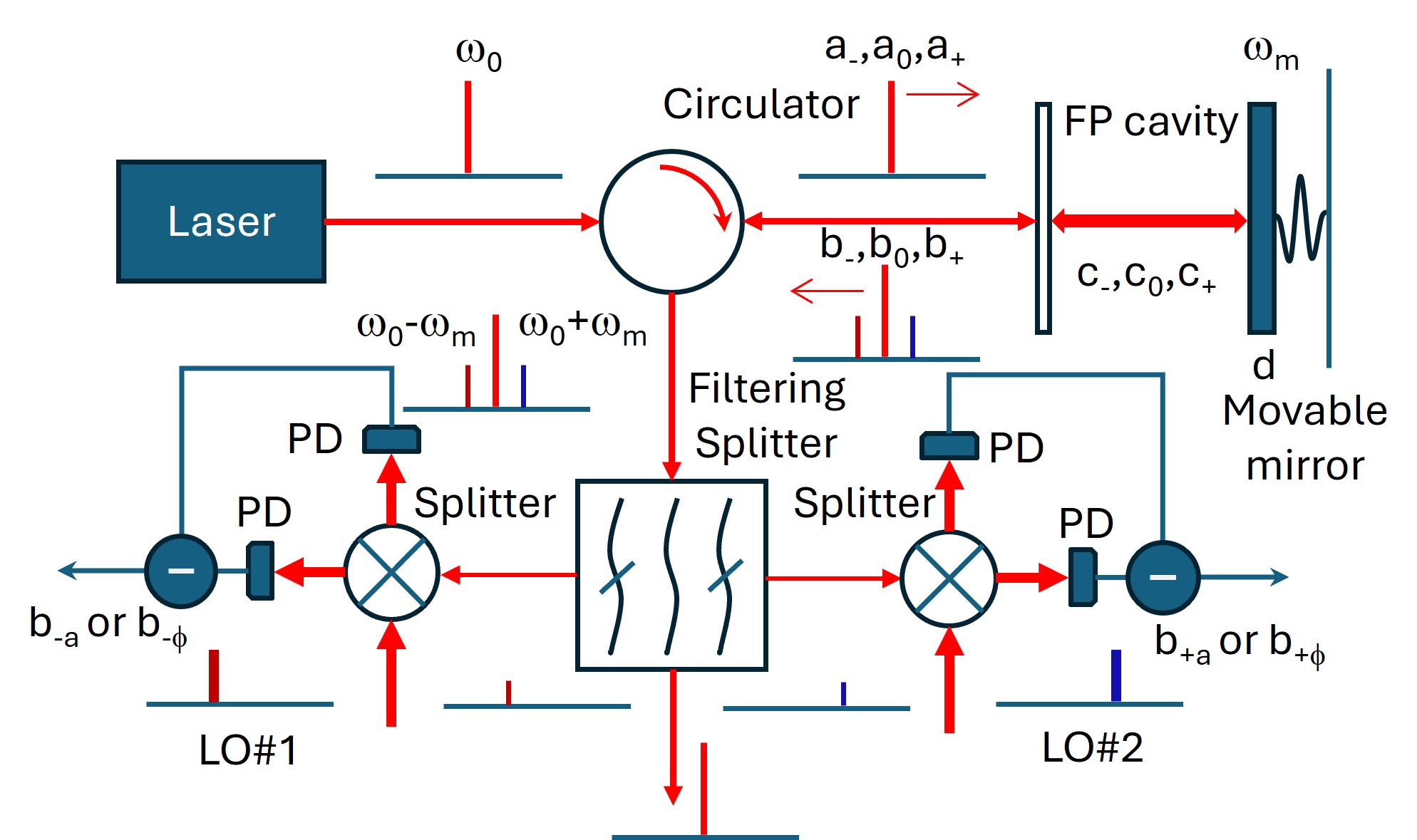}
 \caption{General schematic. Continuous wave classical pump acts on the central mode $\omega_0$ of the optical triplet of a Fabry-Perot cavity, sidebands of which $\omega_\pm=\omega_0\pm \omega_m$ are shifted by $\omega_m$ from central frequency. The sidebands are equidistant and can belong to different mode families. Output radiation from the sidebands is detected separately by balanced homodyne detectors consisting of splitters and pairs of photodiodes (PDs) along with two separate classical local oscillators (LO). In this way one needs to detect either amplitude ($b_{\pm a}$) or phase ($b_{\pm \phi}$) quadrature for each sideband and then combine them to create a collective variable. }\label{scheme}
\end{figure}
Here $\hslash$ is the Planck constant. The term $H_0$ represents the free Hamiltonian describing the energies of the optical and the mechanical oscillator modes. The operators $\hat c_0$, $\hat c^\dag_0$, $\hat c_\pm$, and $\hat c^\dag_\pm$ denote the annihilation and creation operators corresponding to the optical modes, while $\hat d$ and $\hat d^\dag$ are the annihilation and creation operators associated with the mechanical mode (see Fig.~\ref{scheme}). The position operator $\hat x$ of the mechanical oscillator is expressed as
\begin{align}
\label{x}
 \hat x = x_0 \left( \hat d + \hat d^\dag \right), \quad 
 x_0 = \sqrt{\frac{\hslash}{2 m \omega_m}},
\end{align}
where $m$ is the effective mass of the oscillator and $\omega_m$ is its mechanical resonance frequency. The interaction term $H_\text{int}$ represents the optomechanical coupling Hamiltonian. In this analysis, we consider a non-symmetric coupling configuration and introduce distinct coupling constants $\eta_\pm \sim x_0 \omega_0 / L$, where $L$ denotes the effective length of the cavity. The asymmetry of the coupling constants come from the different geometry of the modes involved into the interaction. Without loss of generality we consider real coupling constants in what follows.

\subsection{Langevin equations}

We denote the input quantum optical field amplitudes by $\hat a_{\pm}$ and $\hat a_0$. Using the Hamiltonian, we derive the corresponding Heisenberg–Langevin equations of motion for the slowly varying intracavity field amplitudes. These equations describe the temporal evolution of the optical and mechanical modes within the resonator:
\begin{subequations}
\label{c}
\begin{align} 
	\label{c0}
\dot {\hat c}_0+\gamma_0 \hat c_0&=\eta_+\hat c_+ \hat d^\dag - \eta_- \hat c_- \hat d
	 +\sqrt{2 \gamma_0}\,\hat a_0 ,\\
\dot {\hat c}_++\gamma_+ \hat c_+&=-\eta_+ \hat c_0 \hat d 
	+\sqrt{2 \gamma_{+}}\,\hat a_+ , \\
\dot { \hat c}_-+\gamma_- \hat c_-&=\eta_-\hat c_0 \hat d^\dag 
	 + \sqrt{2 \gamma_{-}}\,\hat a_-\\
\dot {\hat d}+\gamma_m \hat d&=\eta_-\hat c_-^\dag \hat c_0  + \eta_+\hat c_0^\dag \hat c_+  +\sqrt{2 \gamma_m}\,\hat q .
\end{align}  
\label{moveq}
\end{subequations} 
The coupling rates for the front mirror of the Fabry-Perot cavity are $\gamma_\pm$ and $\gamma_0$. Operator $\hat q$ corresponds to the stochastic Langevin force acting on the mechanical oscillator. 
The input quantum  (shot) noise operators $\hat a_\pm$ and the thermal force operator $\hat q$ are characterized by the following commutation and correlation relations:
\begin{subequations}
 \label{commT}	
\begin{align}
	 \label{commA}
	\left[\hat a_\pm(t), \hat a_\pm^\dag(t')\right] &=
	\left\langle\hat a_\pm(t)\, \hat a_\pm^\dag(t')\right\rangle = \delta(t-t'),\\
	\left[\hat q(t), \hat q^\dag(t')\right] &=   \delta(t-t'),\\
	\label{corrQ}
	\left\langle\hat q(t) \hat q^\dag(t')\right\rangle &= (n_T +1)\, \delta(t-t'),\\
	& n_T= \left ( e^{\hslash \omega_m/\kappa_BT} -1 \right )^{-1}. \nonumber
\end{align}
\end{subequations}
Here, $\langle \dots \rangle$ denotes ensemble averaging. The parameter $n_T$ is the thermal occupation number of the mechanical oscillator, $\kappa_B$ is the Boltzmann constant, and $T$ is the ambient temperature. The above relations assume Markovian reservoirs and describe the quantum and thermal noise entering the system through both optical and mechanical channels.

The coupling between the intracavity fields and the external optical modes is governed by the standard input–output boundary conditions:
\begin{align}
\label{outputT}
\hat b_\pm= -\hat a_\pm + \sqrt{2\gamma_{\pm}} \hat c_\pm,
\end{align}
where $\hat b_\pm$ are the output field operators corresponding to the reflected  optical waves.

To facilitate further analysis, it is convenient to separate the steady-state (mean) components of the field amplitudes at the carrier frequency $\omega_0$ from their small quantum fluctuations. Denoting the mean values by capital letters and the fluctuations by lowercase letters, we write
\begin{align}
\label{expA}
\hat c_0 & \Rightarrow C_0 + c_0 ,
\end{align}
where $C_0$ represents the steady-state expectation value of the intracavity field amplitude in the central optical mode, while $c_0$ accounts for its quantum fluctuations. The assumption $|C_0|^2 \gg \langle c_0^\dag c_0 \rangle$ ensures that the fluctuations are small compared to the mean field. Analogous decompositions can be introduced for the sideband optical modes and the mechanical mode. The amplitudes are normalized such that $\hslash \omega_0 |A_0|^2$ corresponds to the optical power of the incident wave \cite{02a1KiLeMaThVyPRD}.

In the following analysis, we assume that the steady-state amplitudes are real quantities:
\begin{align}
\label{real}
A_0=A_0^*,\quad C_0= C_0^*=\sqrt{\frac{2}{\gamma_0}}A_0.
\end{align}
This choice simplifies the analytical treatment and does not affect the generality of the results, as any phase factors can be absorbed into the definitions of the optical field operators.

\subsection{Output fluctuations}

To analyze the system dynamics, the differential equations are solved using the Fourier transform technique. In this treatment, we neglect the influence of the initial conditions and restrict our consideration to the stimulated response of the operators. The Fourier transform of a representative operator, such as $\hat a_\pm$, is introduced as
\begin{align}
\label{apmFT}
\hat a_\pm (t) &= \int_{-\infty}^\infty a_\pm(\Omega)\, e^{-i\Omega t}\, \frac{d\Omega}{2\pi}.
\end{align}
The Fourier components $a_\pm(\Omega)$ satisfy the standard bosonic commutation and correlation relations,
\begin{subequations}
\begin{align}
 \label{comm1}
  \left[a_\pm(\Omega), a_\pm^\dag(\Omega')\right] &= 2\pi\, \delta(\Omega-\Omega'),\\
  \label{corr1}
  \left\langle a_\pm(\Omega) a_\pm^\dag(\Omega')\right\rangle &= 2\pi\, \delta(\Omega-\Omega'),
\end{align}
\end{subequations}
and analogous expressions apply to the remaining noise operators.

Substituting Eqs.~\eqref{expA} and \eqref{real} into the equations of motion \eqref{moveq}, retaining only the terms linear in the small perturbations, and applying the Fourier transform to the resulting expressions yield the following coupled algebraic equations:
\begin{subequations}
	\label{set1}
	\begin{align}
		(\gamma_+ -i\Omega)c_+(\Omega)&=-\eta_+ C_0  d(\Omega) +\sqrt{2 \gamma_{+}}\, a_+ (\Omega) \\
		(\gamma_- -i\Omega) c_-(\Omega)&=\eta_- C_0  d^\dag(-\Omega)   + \sqrt{2 \gamma_{-}}\, a_-(\Omega)\\
		(\gamma_m -i\Omega) d(\Omega)&= C_0\big[\eta_-  c_-^\dag(-\Omega) +  \eta_+ c_+(\Omega)\big]+\\
		&\qquad  +\sqrt{2 \gamma_m}\,\hat q(\Omega) .\nonumber
	\end{align}
\end{subequations}
It follows from this set that the fluctuations of the optical mode near the central frequency $\omega_0$ do not couple to the sideband modes at $\omega_\pm$. Consequently, the equation governing the field component \eqref{c0} is dynamically independent and is omitted from further consideration.

We focus on the output field amplitudes $\hat b_{\pm}$, which represent the measurable quantities of the system. In practice, these quantities can be accessed experimentally using a balanced homodyne detection scheme (Fig.~\ref{scheme}). To facilitate the analysis, we define the amplitude and phase quadratures of the field operators as follows:
\begin{subequations}
\label{quadDef}
\begin{align}
a_{\pm a} &= \frac{a_\pm (\Omega) +a_\pm ^\dag(-\Omega)}{\sqrt 2},\\
a_{\pm \phi} &= \frac{a_\pm (\Omega) -a_\pm ^\dag(-\Omega)}{i\sqrt 2}.
\end{align}
\end{subequations}
Using Eqs.~\eqref{outputT}, \eqref{set1}, and \eqref{quadDef}, we obtain the following expressions for the amplitude and phase quadratures of the output fields $\hat b_{\pm}$:
\begin{subequations}
\label{bpmVy}
\begin{align}
b_{\pm a} &= \left(\xi_{\pm} \mp A_{\pm}\right)a_{{\pm}a} \mp B a_{\mp a} \mp F_\pm\sqrt{2\gamma_m} q_a,\\
b_{\pm \phi} &= \left(\xi_{\pm} \mp A_\pm\right)a_{\pm\phi} \pm B a_{\mp \phi} - F_\pm \sqrt{2\gamma_m}q_{\phi},
\end{align}
\end{subequations}
where the coefficients are defined as
\begin{subequations}
\label{bpm1Vy}
\begin{align}
A_\pm&= \frac{2\gamma_{\pm} \Gamma_{\pm}}{\left(\gamma_{\pm} - i\Omega\right)\left(\Gamma_m - i\Omega\right)},\quad \xi_\pm=\frac{\gamma_{\pm} + i\Omega}{\gamma_{\pm} - i\Omega},\\
\mathcal A_+&=\xi_+ -A_{+} = \xi_+ \left(\frac{\gamma_m - \Gamma_+^* -\Gamma_- -i\Omega }{\Gamma_m -i\Omega} \right),\\
\mathcal A_-&=\xi_- +A_{-} = \xi_-\left(\frac{\gamma_m + \Gamma_+ + \Gamma_-^* -i\Omega }{\Gamma_m -i\Omega} \right),\\
B &= \frac{2}{\Gamma_m - i\Omega}\sqrt \frac{\gamma_+\gamma_-\Gamma_+\Gamma_-}{\left(\gamma_+ - i\Omega\right)\left(\gamma_- - i\Omega\right)},\\
F_\pm &= \frac{1}{\Gamma_m - i\Omega} \sqrt\frac{2\gamma_{\pm}\Gamma_{\pm}}{\gamma_{\pm} - i\Omega},\\
\label{Gammapm}
\Gamma_{\pm} &= \frac{\eta_{\pm}^2C_0^2}{\gamma_{\pm}-i\Omega}, \quad
\Gamma_m = \gamma_m + \Gamma_+-\Gamma_-,\\
\Gamma_{m0} &= \Gamma_m|_{\Omega=0}=\gamma_m +G_+-G_-,\quad G_\pm =\frac{\eta_\pm^2 C_0^2}{\gamma_\pm}
\end{align}
\end{subequations}
Here, $\Gamma_{m0}$ represents the effective bandwidth of the mechanical oscillator, modified by the optical back action (nonlinear optical damping).

In the limiting case where $\Omega \rightarrow 0$, Eqs.~\eqref{bpmVy} can be simplified to
\begin{subequations}
\label{bpmVy2}
\begin{align}
b_{\pm a} &= \mathcal A_{\pm}a_{{\pm}a} \mp B a_{\mp a} \mp F_\pm\sqrt{2\gamma_m} q_a,\\
b_{\pm \phi} &= \mathcal A_\pm a_{\pm\phi} \pm B a_{\mp \phi} - F_\pm \sqrt{2\gamma_m}q_{\phi},\\
\mathcal A_\pm &\simeq \frac{\gamma_m \mp G_+ \mp G_- }{\gamma_m + G_+-G_- } ,\quad
B \simeq \frac{2\sqrt{G_+G_-}}{\gamma_m + G_+-G_-},\\
F_\pm & \simeq \frac{\sqrt{2 G_{\pm}}}{\gamma_m + G_+-G_-}.
\end{align}
\end{subequations}
The corresponding spectral density $S_{b_{\pm a}}$ of amplitude quadrature can  be estimated as:
\begin{align}
\nonumber S_{b_\pm a} &\simeq \frac{ (G_+ +G_- \mp \gamma_m)^2 +4G_+G_-}{(\gamma_m + G_+-G_-)^2 } + \\ &+ \frac{4 G_{\pm}\gamma_m(2n_T+1)}{(\gamma_m +G_+-G_-)^2}.
\end{align}
It follows directly from this expression that $S_{b_{\pm a}}> 1$, indicating that each quadrature, when measured individually, exhibits a noise level exceeding the shot-noise limit. Same result is valid for $S_{b_{\pm \phi}}> 1$.

\subsection{Continuous variable entanglement}

The entanglement criterion based on the sum of dispersions, commonly referred to as the Duan–Simon entanglement criterion \cite{duan2000inseparability,simon2000peres}, provides a rigorous measure of continuous-variable entanglement by evaluating the total dispersion of a pair of operators of EPR type. For the analysis of our optomechanical system, we employ this criterion using a slightly modified set of EPR-like operators:
\begin{subequations}
	\label{EPR2}
	\begin{align}
		&\hat{u} = \hat{x}_+ e^{i\phi_+}\cos\theta + \hat{x}_- e^{i\phi_-}\sin\theta, \\
		&\hat{v} = \hat{p}_+ e^{i\phi_+}\cos\theta - \hat{p}_- e^{i\phi_-}\sin\theta.
	\end{align} 
\end{subequations}

In this notation, the operators 
$\hat x_{\pm}$ and $\hat p_{\pm}$ denote the quadrature operators of the two modes $\omega_+$ and $\omega_-$, while the parameters $\theta$ and $\phi_{\pm}$ allow optimization of the relative weighting and phase of the quadratures. According to this criterion, a two-mode state is entangled if the sum of the dispersions of the operators 
$\hat u$ and $\hat v$ satisfies:
\begin{align}
	\label{EntanglCriterion}
	\left\langle(\Delta\hat{u})^2 \right\rangle + \left\langle(\Delta\hat{v})^2 \right\rangle < 1.
\end{align}

Experimentally, the quadratures of the output fields $b_{\pm}$ can be measured using a balanced homodyne detector, and their Fourier-domain amplitude and phase quadratures are given by Eq. \eqref{bpmVy}. To verify whether the modes $b_{\pm}$ are quantum-entangled, it is necessary to check that the inequality \eqref{EntanglCriterion} holds. For many practical purposes, it is convenient to reformulate the criterion in terms of spectral densities rather than total dispersions. In this representation, the criterion can be expressed as:
\begin{align}
	\label{EntanglCriterionS}
S_{\beta_{a+}} + S_{\beta_{\phi-}} < 2.
\end{align}
for one-sided spectral densities $S_{\beta_{a+}}$ and $S_{\beta_{\phi-}}$ of EPR-like variables
\begin{subequations}
	\label{EPR3}
	\begin{align}
		&\beta_{a+}(\Omega) = b_{+a}(\Omega) e^{i\phi_+}\cos\theta + b_{-a}(\Omega) e^{i\phi_-}\sin\theta,\\
		&\beta_{\phi-}(\Omega) = b_{+\phi}(\Omega) e^{i\phi_+}\cos\theta - b_{-\phi}(\Omega) e^{i\phi_-}\sin\theta.
	\end{align} 
\end{subequations}
The quadrature operators satisfy the canonical commutation relations:
\begin{align}\left[b_{+,- a}(\Omega), b_{+,-\phi}(\Omega')\right] = 2\pi \delta(\Omega - \Omega').
\end{align}
If inequality \eqref{EntanglCriterionS} is satisfied for certain spectral frequencies $\Omega$, this indicates that the two-mode system is in a quantum-entangled state. The spectral form of the criterion allows a direct analysis of frequency-dependent entanglement, which is particularly relevant for optomechanical systems where the quadratures $b_{+a}$, $b_{-a}$, $b_{+\phi}$, and $b_{-\phi}$, are naturally measured in the Fourier domain.

\subsection{Entanglement detection}

We propose to measure the amplitude quadratures $b_{+a}$ and $b_{-a}$ individually and then form their weighted linear combinations  $\beta_{a+}$ and $\beta_{\phi -}$ \eqref{EPR3} to reveal the presence of entanglement between the two optical modes. To satisfy the entanglement condition given by \eqref{EntanglCriterionS},  it is necessary to find the sum of the one-sided spectral densities of the EPR variables $\beta_{a+}$ and $\beta_{\phi -}$, but a straightforward calculation shows that for our system, the spectral densities of these two combinations are equal:
\begin{equation}
S_{\beta_{a+}} = S_{\beta_{\phi-}}.
\end{equation}
Therefore, to characterize the entangled properties of a system, it is sufficient to estimate only $S_{\beta_{a+}}$ relative to unity.

Assuming that the input fluctuations $a_{\pm a}$ correspond to the quantum vacuum noise, the spectral density of $\beta_{a+}$ can be derived from Eqs.~(\ref{bpmVy}, \ref{EPR3}) and expressed as follows:
\begin{subequations}
\label{SE}
\begin{align}
S_{\beta_{a+}} &= S_q +S_T,\\
\nonumber S_q &=\left|\mathcal A_{+}e^{i\frac{\phi}{2}}\cos\theta +Be^{-i\frac{\phi}{2}}\sin\theta\frac{}{}\right|^2+ \\
&+ \left|\mathcal A_{-}e^{-i\frac{\phi}{2}}\sin\theta -Be^{i\frac{\phi}{2}}\cos\theta\frac{}{}\right|^2, \\
\nonumber S_T &=2\gamma_m\ \left|F_-e^{-i\frac{\phi}{2}}\sin\theta - F_+e^{i\frac{\phi}{2}}\cos\theta \right|^2 \times \\ & \times \left(2n_T + 1 \right),\\
\label{phi}
\phi &= \phi_+ - \phi_-.
\end{align}
\end{subequations}
where $S_q$ represents the contribution from quantum fluctuations and $S_T$ accounts for the thermal noise of the mechanical oscillator.  The coefficients $\theta $ and $\phi$ should be selected so that the spectral density $S_{\beta_{a+}}$ takes the smallest possible value at all spectral frequencies.
\begin{figure}
\includegraphics[width=0.45\textwidth]{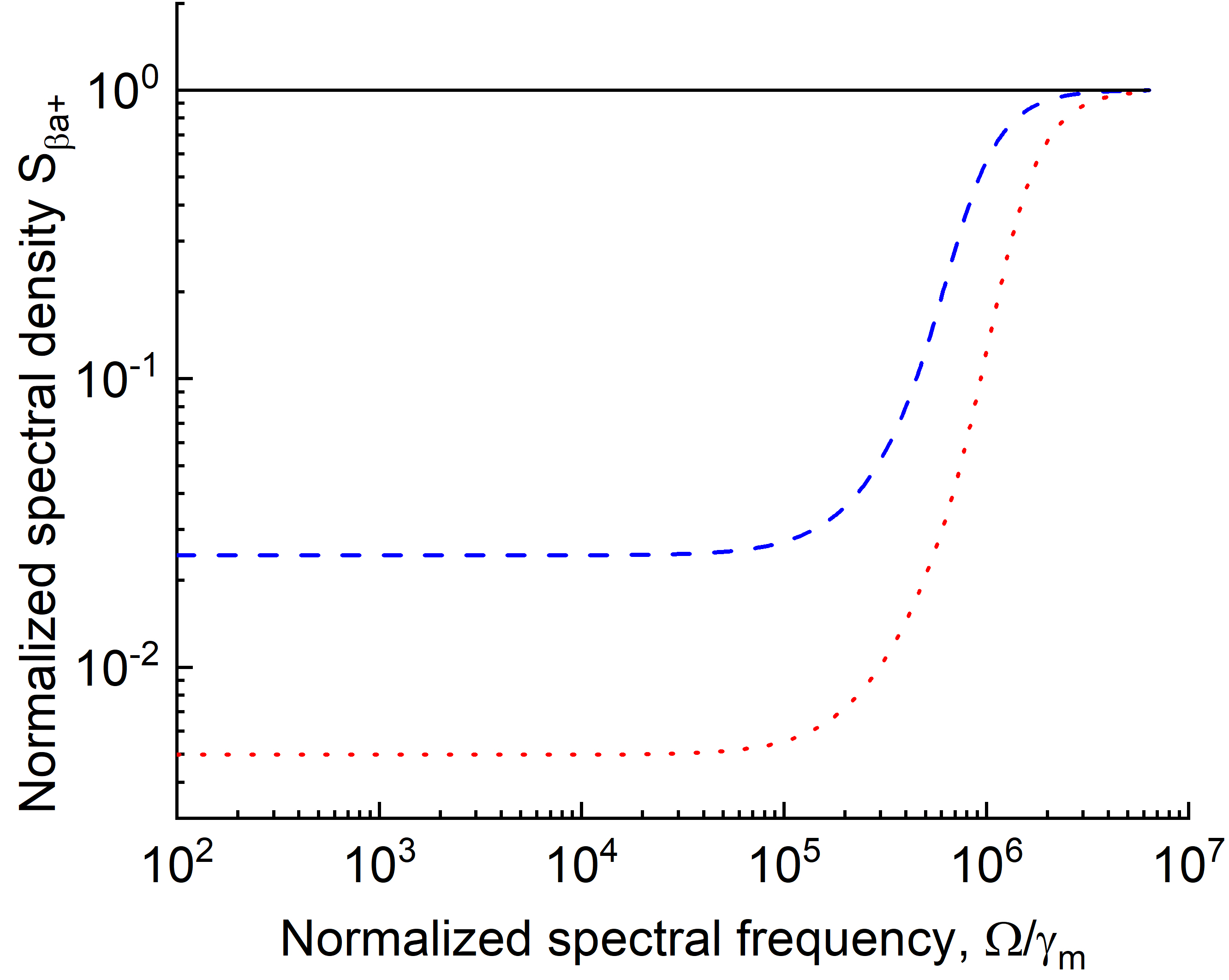}
    \caption{Spectral density $S_{\beta_{a+}} $ \eqref{SE} as function of spectral frequency $\Omega$.  Solid black line stands for SQL. Dashed blue line stands for the input power $P_{in}$ and other parameters from Table \ref{table1}, dotted red line corresponds to five time larger ($\times 5$) $P_{in}$. Parameters $\theta $ and $\phi$ are selected in the optimal way, $G_+ = G_-$.}\label{fig1}
\end{figure}

The minimum of the spectral density $S_{\beta_{a+}}$ is achieved when
\begin{align}
\label{simple}
\Omega &=0,\ \Rightarrow\ \xi_\pm=1,\quad \Gamma_\pm =\Gamma_\pm^*= G_\pm.
\end{align}
In case $\Gamma_m\ll G_{\pm}$, this minimum has the following form
\begin{subequations}
\label{SE2}
\begin{align}
& \nonumber S_{\beta_{a+}} |_{\Omega=0}^\text{min} \simeq \frac{\gamma_m (2n_T + 1)}{2G +\gamma_m (2n_T +1)} + \\ &+ \frac{\gamma_m\Gamma_m \left(G + (2n_T +1)\gamma_m\right)}{2G\left(2G + (2n_T +1)\gamma_m\right)^2}\times \\&\left[\frac{\Gamma_m\left(G + (2n_T +1)\gamma_m\right)}{\gamma_m\left(2G + (2n_T +1)\gamma_m\right)} - 1\right], \\
&G = \frac{G_+ + G_-}{2}.
\end{align}
\end{subequations}

Demanding $G \gg \gamma_m$ we arrive at a simplified expression
\begin{equation}
   S_{\beta_{a+}} |_{\Omega=0}^\text{min} \simeq \frac{\gamma_m (2n_T +1)}{2G+\gamma_m (2n_T +1)} \label{minapprox}
\end{equation}

This equation shows that to obtain good entanglement it is sufficient to require that $G \gg \gamma_m n_T$.

Moreover, the maximum entanglement will be achieved when $\Gamma_m \rightarrow 0$,  despite the strong optomechanical heating of the mechanical oscillator. The average number of phonons in the mechanical subsystem taking into account optomechanical interaction increases in the vicinity of the oscillation threshold 
\begin{align}
N_T = \left<d^{\dagger} d\right> &= \frac{\gamma_m (\gamma + \Gamma_m)}{\Gamma_m(\gamma + \gamma_m)}n_T + \frac{\gamma G_-}{\Gamma_m(\gamma + \gamma_m)}.
\end{align}
Here and throughout, we assume that $\gamma_+ = \gamma_- = \gamma$.

\begin{figure}
\includegraphics[width=0.45\textwidth]{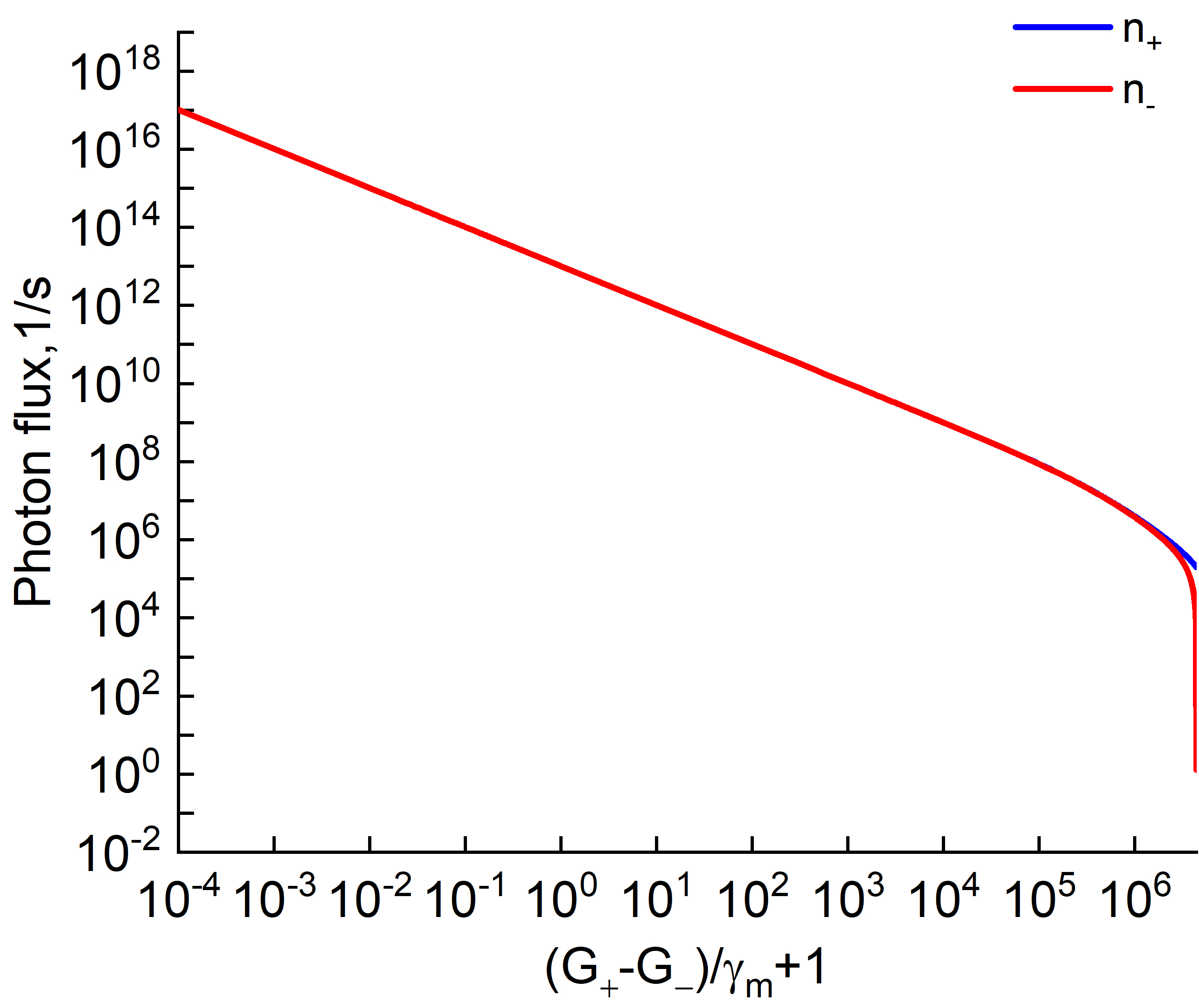}
    \caption{Dependence of the photon fluxes \eqref{flux1} of the $\omega_{\pm}$ modes on the asymmetry of the optomechanical interaction $G_+ - G_-$ at a fixed total power of the optomechanical interaction $(G_+ + G_-)/2 = 2.2\times 10^6$. The blue curve represents the flux $2\gamma n_+$, and the red curve represents $2\gamma n_-$.}  \label{flux}
\end{figure}

The flux of photons leaving the Stokes and anti-Stokes modes is
\begin{subequations}
\label{flux1}
\begin{align}
\langle b_\pm^\dagger b_\pm \rangle &= 2\gamma \langle c_\pm^\dagger c_\pm \rangle = 2\gamma n_{\pm}, \\
\nonumber n_+ &= \frac{
\gamma_m n_T G_+
}{
\Gamma_m(\gamma+\gamma_m)
} + \\ &+
\frac{
G_+G_-(2\gamma+\gamma_m)
}{
\Gamma_m(\gamma+\gamma_m)(\Gamma_m+2\gamma+\gamma_m)
},\\
\nonumber n_- &= \frac{
\gamma_m n_TG_-
}{
\Gamma_m(\gamma+\gamma_m)
} + \\ &+
\frac{G_-\left[
2\Gamma_m(\gamma+\gamma_m)
+
G_-(2\gamma+\gamma_m)
\right]
}{
\Gamma_m(\gamma+\gamma_m)(\Gamma_m+2\gamma+\gamma_m)
}.
\end{align}
\end{subequations}
Fig.~\ref{flux} shows the dependence of the output photon fluxes in the $\omega_{\pm}$ modes on the asymmetry of the optomechanical interaction, $G_+ - G_-$, for a fixed average optomechanical coupling strength $(G_+ + G_-)/2 = 2.2\times10^6$. As the system approaches the generation threshold, $\Gamma_m \rightarrow 0$, the photon fluxes of both modes increase sharply. In the opposite limit, $G_- \rightarrow 0$, the optomechanical interaction associated with the $\omega_-$ mode vanishes, resulting in $2\gamma n_- \rightarrow 0$.

It is evident from Fig.~\ref{flux} that the flux $2\gamma n_-$ remains smaller than $2\gamma n_+$ over the entire range of asymmetry parameters considered. Since each entangled photon pair contributes one photon to each sideband mode, the pair-generation rate is bounded by the smaller of the two photon fluxes. Therefore, the quantity $2\gamma n_-$ provides an upper estimate of the flux of entangled photon pairs generated by the optomechanical interaction.

It is important to mention that the number of pairs increases when the system approaches instability threshold. Our model breaks when the predicted number of generated pairs becomes comparable with the number of the input pump photons. For the selected numerical values the pump photon flux is $5\times 10^{17}$~s$^{-1}$. Therefore, the model needs adjustments when $(G_+-G_-)/\gamma_m+1$ approaches $10^{-4}$. 
\begin{figure}
\includegraphics[width=0.45\textwidth]{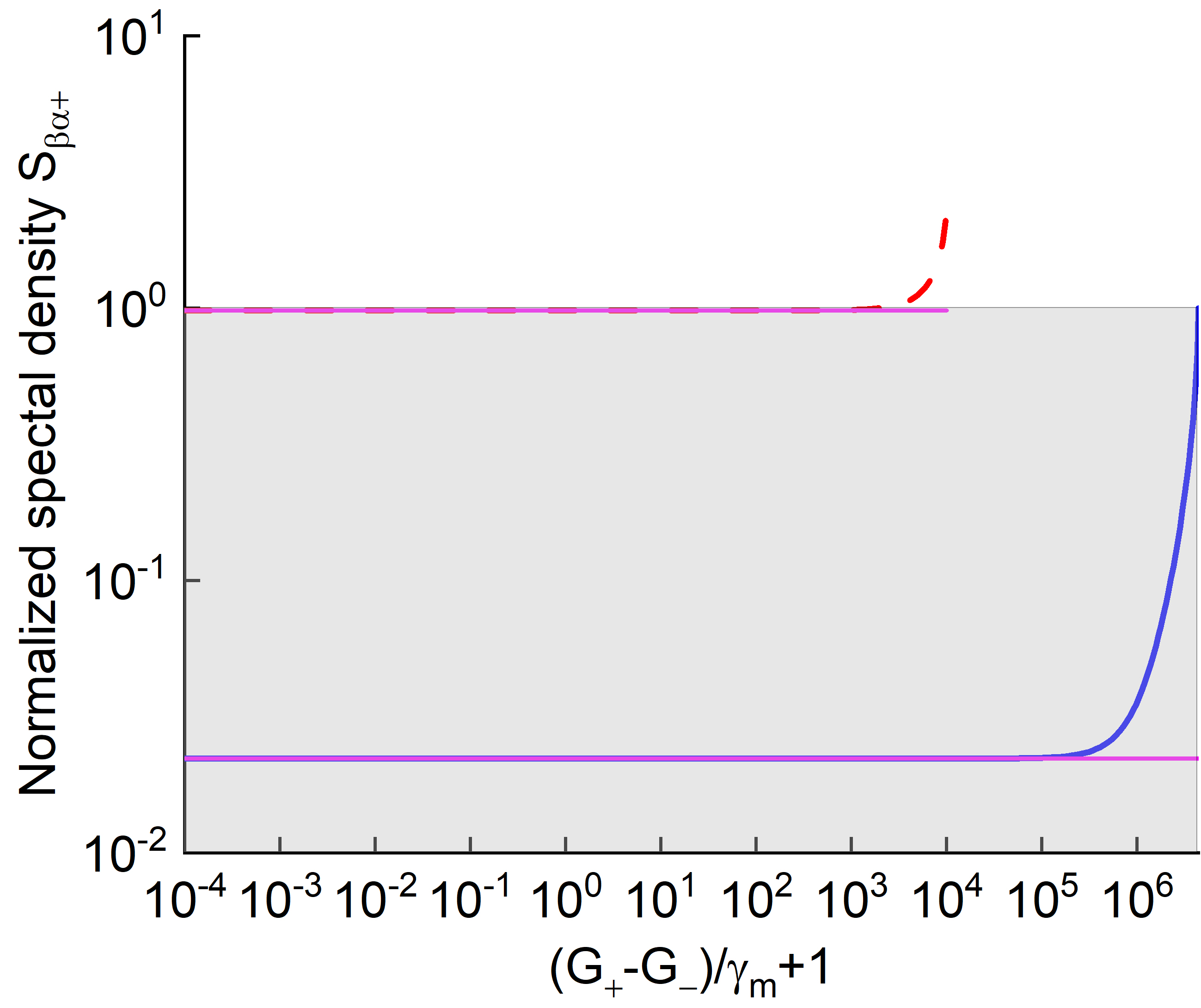}
    \caption{Dependence of the normalized spectral density of the combined quadrature component on the asymmetry of the optomechanical interaction $G_+ - G_-$. The solid blue line corresponds to $G=2.2\times 10^6 \gamma_m$ and dashed red line -- to $G=10^3 \gamma_m$. The other parameters utilized in plotting blue and red lines correspond to those of Figure~\ref{fig1}. The magenta lines are plotted using Eq.~(\ref{minapprox}). The output signals are entangled if the corresponding curves overlap with the gray area area of the plot. }  \label{entang}
\end{figure}

Figure~\ref{fig1} shows the dependence of the minimal spectral density $S_{\beta_{a+}} $ obtained with complex parameters $\theta $ and $\phi$ optimally selected in post-processing. The parameters used in the simulations are summarized in Table~\ref{table1}. The entanglement bandwidth does not depend on the entanglement level and is entirely determined by the bandwidth of the optical resonator.  
\begin{align}
    \Delta \Omega_{ent} \simeq \gamma.
\end{align}
\begin{table}[b]
	\caption{Parameters of the silicon nitride membrane-based mechanical oscillator and the optical cavity employed in the numerical estimates.} \label{table1}
	\begin{tabular}{||c | c | c||}
		\hline
		\multicolumn{3}{||c||}{Membrane} \\
		\hline
		Mass, $m$ & $5\cdot 10^{-11}$ &  kg\\
		Frequency, $\omega_m/2\pi$& $2 \cdot 10^6$ &  Hz\\
		Quality factor $Q=\frac{\omega_m}{2\gamma_m}$ & $0.6\cdot10^7$ & \\
		Temperature, $T$ & 10 & K$^\circ$ \\
		Thermal phonons number, $n_T$ & $1.04\cdot 10^5$ &\\
		% Time of signal force $\tau=30 \cdot\frac{2\pi}{\omega_m}$ & $0.84\cdot 10^{-3}$ & sec \\
		\hline
		\multicolumn{3}{||c||}{Cavity} \\
		\hline
        	Effective
length of the cavity, L & 10 & cm \\
		Bandwidth, $\gamma $ &  $3\cdot 10^{5}$ & s$^{-1}$\\
		% $\gamma_e,\ \gamma_{e\pm}$&  2.3 $10^{3}$ & s$^{-1}$\\
		% Couplings $\eta_\pm$ & $\frac{\omega_0x_0}{L}(1\pm 3\%)$ & \\
		Wave length, $\lambda= 2\pi c/\omega_0$ & $1.55\cdot 10^{-6}$  & m \\ 
		Input power $P_{in} $ & $10^{-2}$ & W \\
        Pump parameter $G=(G_++G_-)/2$ & $2.2\cdot 10^6$ & s$^{-1}$\\
		\hline
	\end{tabular}
\end{table}

Figure~\ref{entang} illustrates the dependence of the entanglement on the system asymmetry. It can be seen that the outputs of the two optical modes remain entangled over a broad range of parameters, demonstrating the robustness of the generated quantum correlations. This robustness makes the entanglement observable in a variety of experimental configurations, including systems involving optical and mechanical modes with different coupling strengths and geometries. Therefore, the effect does not rely on fine tuning of the system parameters and can be expected to persist in realistic optomechanical implementations. 
\begin{figure}
\includegraphics[width=0.45\textwidth]{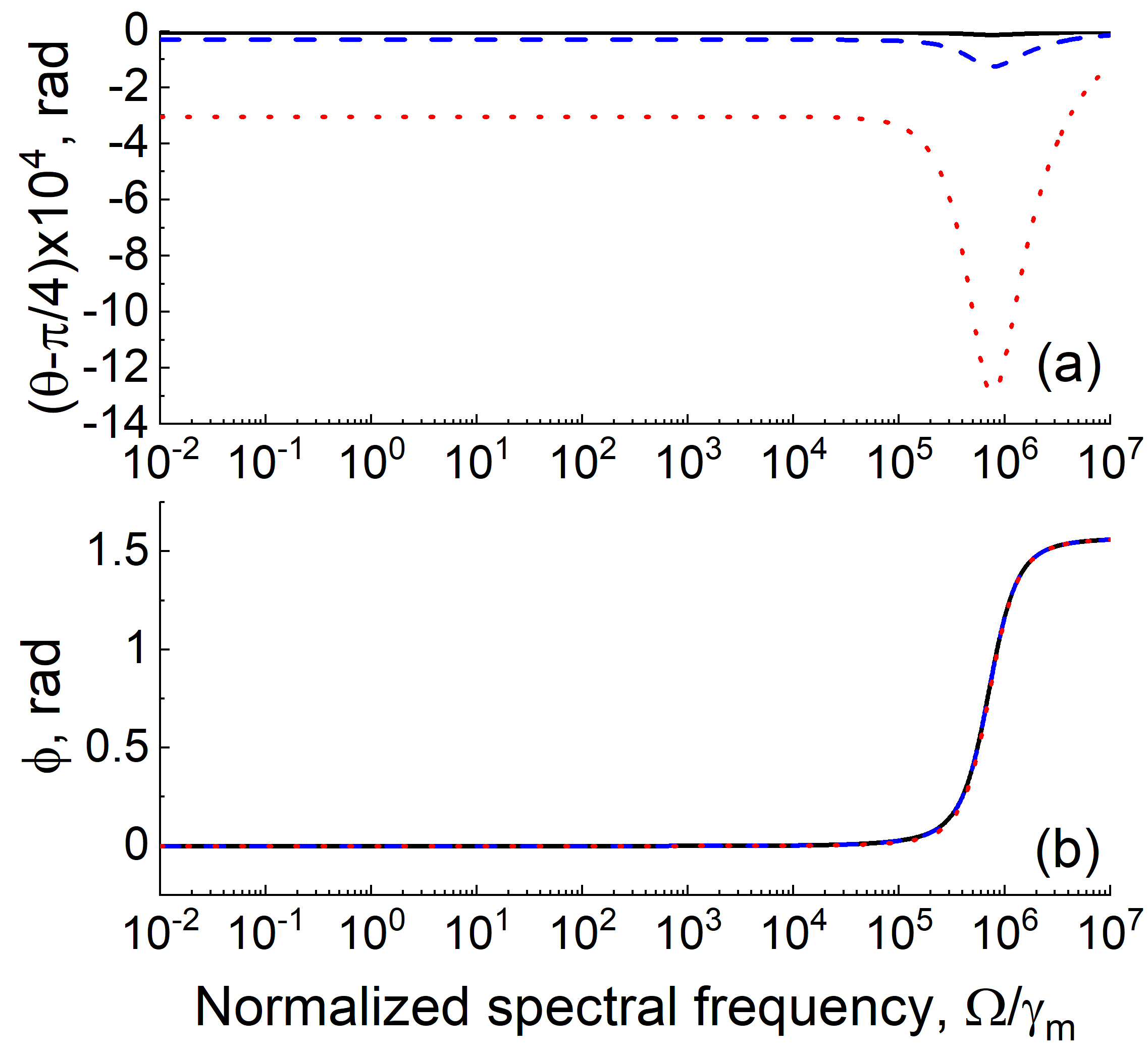}
    \caption{Optimal angles $\theta(\Omega)$ and $\phi(\Omega)$  defined in \eqref{phi} as functions of spectral frequency $\Omega$. The curve correspond to parameters $G=(G_++G_-)/2=2\times 10^6\, \gamma_m$ and $G_+-G_- = 10^3\gamma_m$ (black solid line), $G_+-G_- = 10^4\gamma_m$ (blue dashed line), $G_+-G_- = 10^5\gamma_m$ (red dotted line). Other parameters are taken from Table~\ref{table1}.}\label{figPhi}
\end{figure}

Plots for Fig.~\ref{fig1} are obtained for optimal angle $\theta$ and phase $\phi$ in \eqref{phi}, which in turn are functions of spectral frequency $\Omega$. Their plots are presented on Fig.~\ref{figPhi}. We see that angle $\theta$ slightly depends on $\Omega$, while phase $\phi$ monotonically increases from zero to about $\pi/2$.

\section{Discussion}

The primary goal of this paper is to provide proof-of-principle evidence for entanglement generation in a resonant optomechanical system interrogated with continuous wave light. Using the parameter set listed in Table~\ref{table1}, we numerically demonstrate that the system can operate in a regime where entanglement is produced reliably. These parameters satisfy the conditions

\begin{subequations}
\begin{align}
&\frac{G}{\gamma_m} \gg n_T > 1;\\ 
&\omega_m \gg \gamma_0,\gamma_{\pm} \gg \gamma_m,
\label{eq:cond3}
\end{align}
\end{subequations}

under which resonantly enhanced opto-mechanical interactions are most effective. Equation \eqref{eq:cond3} ensures, respectively, (i) sufficiently strong and well-resolved sideband coupling and dominance of coherent opto-mechanical rates over thermal decoherence, (ii) operation deep in the resolved-sideband regime.

It is worth noting, however, that although this parameter set demonstrates the feasibility of generating entanglement, additional optimization will be required for specific experimental implementations. Such optimization, aimed at maximizing the entanglement strength, robustness, and detection efficiency, will be discussed elsewhere.

\subsection{Entanglement and oscillator temperature}
\begin{figure}
\includegraphics[width=0.45\textwidth]{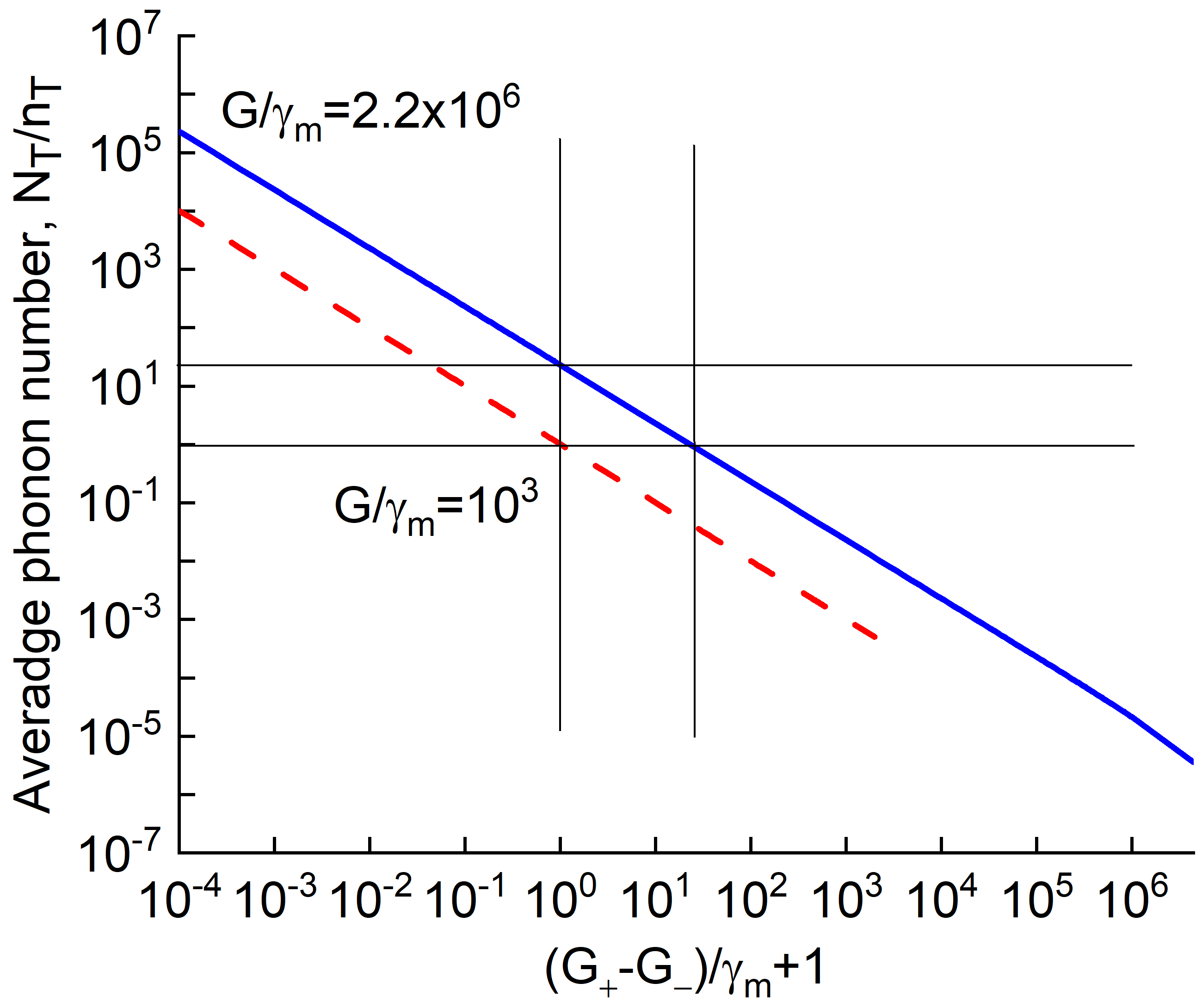}
    \caption{Dependence of the average number of phonons $N_T$ on the asymmetry of the optomechanical interaction $G_+ - G_-$. It is interesting to note that for the case of larger $G$ the number of thermal quanta in the system exceeds $n_T$ even when $G_+=G_-$. The Nyquist thermal number is reached when opto-mechanical cooling is involved. Still, the entanglement level does not depend on the number of thermal phonons significantly till the point when the mechanical system is optically cooled to nearly zero temperature. The parameters used in the simulations are summarized in Table~\ref{table1}.}\label{phonons}
\end{figure}

As the system tends to instability, the quantum entanglement of optical modes does not change (Figure~\ref{entang}), while the average number of phonons in the mechanical subsystem increases sharply (Figure~\ref{phonons}). This means that optical heating of a mechanical oscillator does not interfere with quantum entanglement.

Depending on the sign of the optical additive $G_+ - G_-$, the effective quality factor of the mechanical oscillator can either increase or decrease \cite{CavityCool}. The  optical damping decreases the mean number $n_T'$ of thermal phonons:
\begin{align}
    n_T'& \simeq \frac{\gamma_m}{|\Gamma_m|}\, n_T
        \simeq \frac{\gamma_m}{G_+-G_-}\, n_T \ll n_T.
\end{align}

It is interesting to observe the entanglement change in the case of opto-mechanical cooling. In this regime, the quality factor of the mechanical oscillator is significantly reduced, which in turn suppresses the number of thermal phonons in the mechanical mode. The significant reduction of the phonon number leads to reduction of the entanglement (Figure~\ref{entang}).

\subsection{Distinction with the parametric entanglement}

It is instructive to compare the performance of the conventional parametric entanglement source with that of the optomechanical source described here. To this end, let us consider a non-degenerate optical parametric amplifier. The amplifier is assumed to be implemented as a Fabry–Perot cavity containing a nonlinear $\chi^{(2)}$ crystal. An electromagnetic pump wave with amplitude $E_0$ and frequency $2\omega_0 = \omega_1 + \omega_2$ propagates through the crystal, where $\omega_1$ and $\omega_2$ are the resonant frequencies supported by the optical cavity.

The temporal evolution of the interacting optical modes can then be described by the following set of coupled equations:
\begin{subequations}
	\label{amplitudes}
	\begin{align}
		&\dot{\hat{c}}_1 +\gamma_1\hat{c}_1= G_{par}\hat{c}_2^{\dagger} +\sqrt{2\gamma_1}\hat{a}_1,\\
		&\dot{\hat{c}}_2+ \gamma_2\hat{c}_2 = G_{par}\hat{c}_1^{\dagger} +\sqrt{2\gamma_2}\hat{a}_2.
	\end{align}   
\end{subequations}
Here $\hat{c}_1$, $\hat{c}_2$ and $\hat{a}_1$, $\hat{a}_2$ are operators of slow amplitudes
of the intracavity and external optical fields (defined identically with \ref{c} and \ref{commT}), respectively. The parameter $G_{par} = \chi^{(2)}E_0$ stands for the parametric gain. The output field is found from expressions identical to \eqref{outputT}.

Let us introduce the amplitude and phase quadratures of the electromagnetic fields:
\begin{align}
	\label{quad}
	a_a &= \frac{a(\Omega)+ a^\dag(-\Omega)}{\sqrt 2},\quad a_{\phi} = \frac{a(\Omega) - a^\dag(-\Omega)}{i\sqrt 2}.
\end{align}
and combinations of the quadratures $g_- = (b_{1a} - b_{2a})/\sqrt{2}$ and $g_+ = (b_{1\phi} + b_{2\phi})/\sqrt{2}$. We find for $g_-$
\begin{align} \nonumber
	g_- &= \frac{(\sqrt{\gamma_1\gamma_2} - G_{par})^2 - i\Omega(\gamma_1 - \gamma_2) + \Omega^2}{(\gamma_1 - i\Omega)(\gamma_2 - i\Omega) - G_{par}^2}\frac{a_{1a}}{\sqrt{2}} - \\ &-  \frac{(\sqrt{\gamma_1\gamma_2} - G_{par})^2 + i\Omega(\gamma_1 - \gamma_2) + \Omega^2}{(\gamma_1 - i\Omega)(\gamma_2 - i\Omega) - G_{par}^2}\frac{a_{2a}}{\sqrt{2}},\\
     \nonumber
    g_+ &= \frac{(\sqrt{\gamma_1\gamma_2} - G_{par})^2 - i\Omega(\gamma_1 - \gamma_2) + \Omega^2}{(\gamma_1 - i\Omega)(\gamma_2 - i\Omega) - G_{par}^2}\frac{a_{1\phi}}{\sqrt{2}} + \\ &+  \frac{(\sqrt{\gamma_1\gamma_2} - G_{par})^2 + i\Omega(\gamma_1 - \gamma_2) + \Omega^2}{(\gamma_1 - i\Omega)(\gamma_2 - i\Omega) - G_{par}^2}\frac{a_{2\phi}}{\sqrt{2}}.
\end{align}
The one-sided power spectral density of the combination $g_\pm$ has the following form:
\begin{align}
	S_{g_\pm}(\Omega) = \frac{\left[\left(\sqrt{\gamma_1\gamma_2} - G_{par}\right)^2 + \Omega^2\right]^2 + \Omega^2\left(\gamma_1-\gamma_2\right)^2}{\left(\gamma_1\gamma_2 - G_{par}^2 - \Omega^2\right)^2 + \Omega^2\left(\gamma_1+\gamma_2\right)^2}.
\end{align}

The minimum of this spectral density, i.e. the maximum entanglement, is achieved at $\Omega \rightarrow 0$. This minimum has the following form:
\begin{align}
	\min S_{g_\pm} = \left(\frac{\sqrt{\gamma_1\gamma_2} - G_{par}}{\sqrt{\gamma_1\gamma_2} + G_{par}}\right)^2.
\end{align}

The bandwidth of the entanglement is
\begin{align}
	\Delta \Omega_{\chi^{(2)}} \simeq \frac{4\gamma_1\gamma_2\sqrt{\min S_{g_\pm}}}{\gamma_1+\gamma_2} \left(1 - \frac{4\gamma_1\gamma_2\sqrt{\min S_{g_\pm}}}{\left(\gamma_1+\gamma_2\right)^2} \right).
\end{align}
One can see, that the larger is the entanglement, the smaller is the bandwidth of the entanglement. Interestingly, defining frequency dependent quadratures, similarly to \eqref{EPR3}, does not result in the modification of the bandwidth.

The bandwidth of the opto-mechanically induced entanglement (see Fig.~\ref{fig1}) is not correlated with the minimum of the spectral density $S_{\beta_{a+}} $, which distinguishes this configuration from the conventional case of entanglement generated by a parametric amplifier. This feature highlights a unique regime of optomechanical entanglement, where broad bandwidth and strong quantum correlation can coexist.

\subsection{Ideal vs synodyne detection}
\begin{figure}[!t]
 \includegraphics[width=0.48\textwidth]{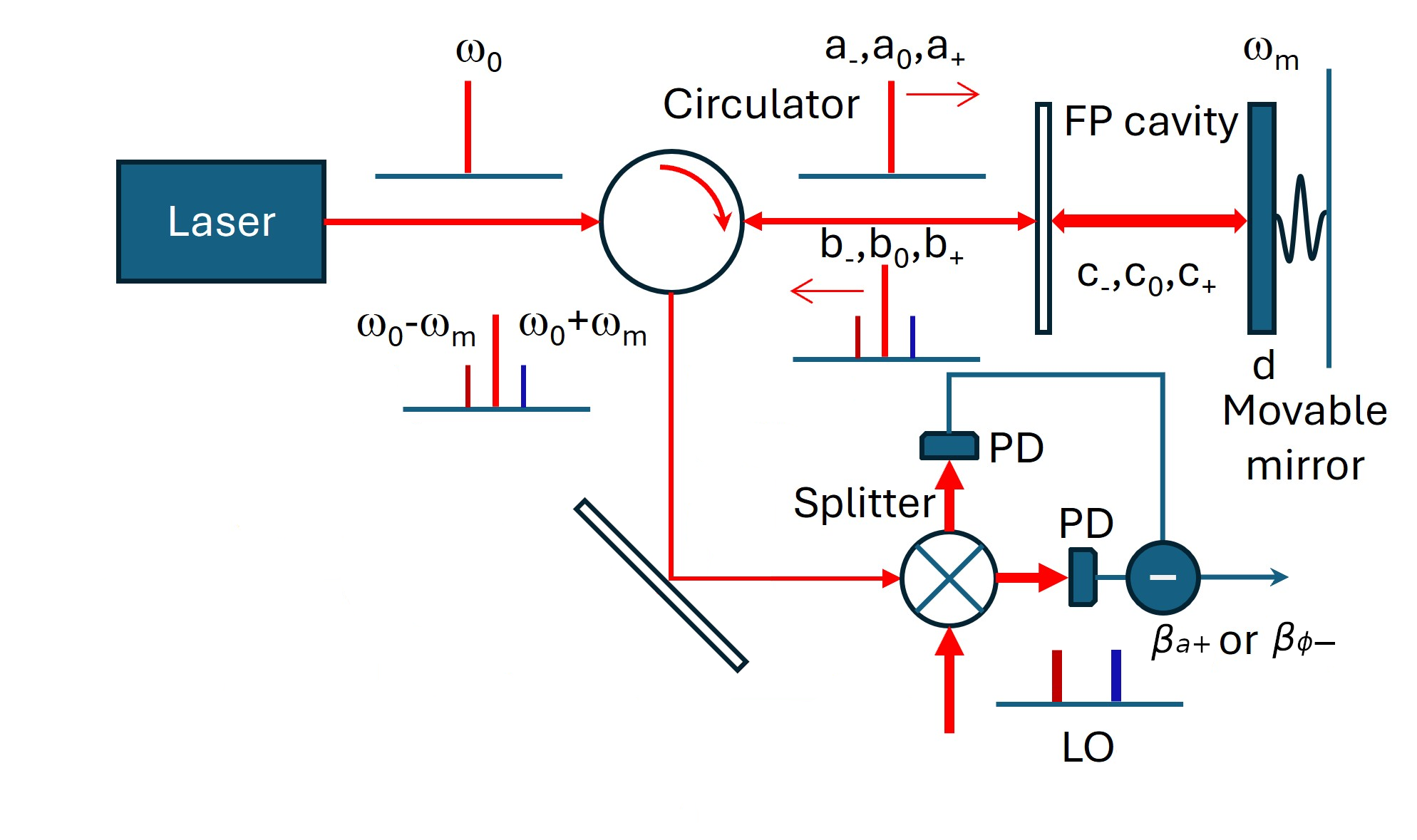}
 \caption{Sinodyne detection schematic: all output radiation, contained output of modes  $\omega_0,\ \omega_\pm$, sends to balanced homodyne detector with a dichromatic local oscillator that contains the two carrier frequencies $\omega_\pm$.}\label{Syno}
\end{figure}

The measurement scheme presented above is hard to realize. To enable post-processing, one must separate the output fields corresponding to the modes $\omega_+$ and $\omega_-$, whose frequencies differ by $2\omega_m$ and also filter out the pump light. This frequency difference is small on the optical scale, making spectral separation experimentally challenging.

Here we propose a simpler method for detecting entanglement based on the so-called synodyne detection technique \cite{Buchmann2016} --- see Fig.~\ref{Syno}. In this approach, the coefficients $\theta $ and $\phi$ cannot be chosen as functions with arbitrary spectral dependence on $\Omega$; instead, they must remain constant. Despite this restriction, the method still enables the observation of entanglement, while offering a much simpler experimental realization.

The local oscillator field $E_{LO}$ of the synodyne detector contains two carrier frequencies:
\begin{align}
E_{LO} &= A_+ e^{-i\omega_+ t} + A_- e^{-i\omega_- t} + \text{conj.}
\end{align}
We are interested in the slowly varying components of the difference photocurrent $I$, that is, in the terms oscillating at frequencies close to zero:
\begin{align} \nonumber
I &\sim \left[A_+^* e^{i\omega_+ t} + A_-^*e^{i\omega_- t}\right]\times\\ 
&\quad \times\left( b_+e^{-i\omega_+ t} + b_0e^{i\omega_0 t}+ b_- e^{-i\omega_- t}\right)+ \text{conj.}
\label{syno}
\end{align}
The rapidly oscillating terms $\sim e^{\pm i\omega_m t}$ (associated with the mode $b_0$ at $\omega_0$), $\sim e^{\pm 2i\omega_m t}$, and higher harmonics should be omitted. Equation \eqref{syno} can be simplified:
\begin{align} 
I &\sim A^*_+ b_+ +A_+ b_+^\dag  + A_-^*b_- + A_-b_-^\dag
\label{syno2}
\end{align}
Equation~\eqref{syno2} demonstrates that when the local oscillator amplitudes are real, $A_\pm=A_\pm^*$, the detection measures a weighted sum $\beta_{a+}$ with constant coefficients $\theta = \pi/4$ and $\phi_+ = \phi_- = 0$. Conversely, when the local oscillator amplitudes are purely imaginary, $A_\pm=-A_\pm^*$, the measured observable corresponds to a weighted sum $\beta_{\phi-}$ with the same coefficients $\theta$ and $\phi_{\pm}$. The spectral density of the sum $\beta_{a+}$ defined in Eq.~\eqref{EPR3} can be written for this case as follows:
\begin{widetext}
\begin{subequations}
\label{Ssyno}
  	\begin{align}
    \label{SsynoA}
  	S_\text{syno} &=  \frac{ 1}{ 2}  \left|\frac{\gamma_m - \left(\sqrt{\Gamma_+^*} -\sqrt{\Gamma_-}\right)^2 -i\Omega  +\Delta \sqrt{\Gamma_+^*\Gamma_-} }{\gamma_m +\Gamma_+-\Gamma_- -i \Omega}  \right|^2 
     + \frac{ 1}{ 2}  \left|\frac{\gamma_m + \left(\sqrt{\Gamma_+} -\sqrt{\Gamma_-^*}\right)^2 -i\Omega  -\Delta^* \sqrt{\Gamma_+\Gamma_-^*} }{\gamma_m +\Gamma_+-\Gamma_- -i \Omega}  \right|^2 + \\
     \label{SsynoB}
  	 + & \gamma_m\left[2n_T +1\right]\left| \frac{-\sqrt{1+ \xi_+}\,\sqrt{\Gamma_+} +\sqrt{1+ \xi_-}\,\sqrt{\Gamma_-}}{\gamma_m +\Gamma_+-\Gamma_- -i \Omega} \right|^2,
    \qquad \Delta = \sqrt{(1+\xi_+^*)(1+\xi_-)} - 2.
  	\end{align}
\end{subequations}
\end{widetext}

The approximate expressions \eqref{SE2} derived for the spectral density of the noise when $\Omega \rightarrow 0$  remain valid for the synodyne case.  However, the synodyne detection may have narrower detection bandwidth if compared with the ideal one. The phase $\phi$ defined in Eq.~\eqref{phi} cannot be optimally chosen in the case of synodyne measurement. Figure~\ref{fig4} shows the spectral density $S_\text{syno}$ for several values of optical damping $\Gamma_{m0}$ using the parameters listed in Table~\ref{table1}. For comparison, the figure also includes the corresponding spectral density $S_{\beta_{a+}}$. It can be seen that the two spectral densities coincide at zero frequency, while their entanglement bandwidths differ. For $\Gamma_{m0}\simeq \gamma_0$, the bandwidths of $S_{\beta_{a+}}$ and $S_\text{syno}$ are similar; however, when $\Gamma_{m0}\ll\gamma_0$, the bandwidth of $S_\text{syno}$ becomes significantly narrower.

Let us estimate the bandwidth of the synodyne detection. The optimal phase in $S_{\beta_{a+}}$ is primarily determined by the mechanical susceptibility, which becomes smoother when $\Gamma_{m0}$ increases due to increase of the optical damping. Consequently, the effective bandwidth of $S_\text{syno}$ increases with opto-mechanical damping, illustrating the trade-off between experimental simplicity and optimal entanglement detection.

To define the measurement bandwidth $\Delta\Omega$ we demand $S_\text{syno}(\Delta \Omega) = 2S_{\beta_{a+}}|_{\Omega=0}^\text{min}$ which results in
\begin{align}
\label{DeltaO}
\Delta\Omega_\text{syno} \simeq 2\Gamma_m\sqrt{S_{\beta_{a+}}|_{\Omega=0}^\text{min}}.
\end{align}

\subsection{Simultaneous detection possibility}

An important feature of the synodyne detection scheme is its ability to access both amplitude and phase EPR-like observables simultaneously within a single measurement channel. Let the amplitudes of the bichromatic local oscillator be written as

\begin{equation}
A_\pm = |A|e^{i\theta_\pm}.
\end{equation}

The slowly varying component of the difference photocurrent can then be expressed as

\begin{equation}
I \propto
b_{+a}\cos\theta_+
-b_{+\phi}\sin\theta_+
+b_{-a}\cos\theta_-
-b_{-\phi}\sin\theta_-.
\end{equation}

Choosing the local-oscillator phases as

\begin{equation}
\theta_+=\frac{\pi}{4},
\qquad
\theta_-=-\frac{\pi}{4},
\end{equation}

one obtains

\begin{equation}
I \propto
(b_{+a}+b_{-a})
-
(b_{+\phi}-b_{-\phi}),
\end{equation}
which corresponds to a simultaneous measurement of the EPR-like amplitude and phase combinations $\beta_{a+}$ and $\beta_{\phi-}$.

Since the optical input fields are assumed to be in coherent states, the amplitude and phase quadratures of the vacuum fluctuations are uncorrelated. As a consequence, the cross-spectral density between $\beta_{a+}$ and $\beta_{\phi-}$ vanishes, and the spectral density of the photocurrent reduces to

\begin{equation}
S_I(\Omega)
=
S_{\beta_{a+}}(\Omega)
+
S_{\beta_{\phi-}}(\Omega).
\end{equation}

Therefore, the quantity entering the Duan--Simon inseparability criterion can be measured directly with a single synodyne detector, without the need for separate measurements of the amplitude and phase EPR observables.

\subsection{Improvement of a ponderomotive sensor}

The configuration of the broadband variational measurement of a force \cite{21a1VyNaMaPRA} resembles the setup analyzed in this work. Namely, the pumping of the principal optical mode $\omega_0$ and detection of the output quadratures of the radiation from modes $\omega_\pm$, which are separated from the main mode by the frequency of the mechanical oscillator, $\omega_\pm = \omega_0 \pm \omega_m$, allows removal of the quantum back action and broadband detection of a classical force, acting on the mechanical oscillator. For instance, one can detect the output amplitude quadratures $j_{\pm a}$ and form their combinations as $\delta_{a+} = \big(j_{+a} + j_{a-}\big)/\sqrt{2}$ and $ \delta_{a+} = \big(j_{+a} - j_{-a}\big)/\sqrt{2}$ to perform the back action evading measurement. In the symmetric case ($\eta_+ = \eta_- = \eta,\ \gamma_+ = \gamma_- = \gamma$) these combinations take the following form:
\begin{subequations}
 \label{betaMSIa}
 \begin{align}
  \delta_{a-}   &= \xi\, \epsilon_{a-},\quad \xi=\frac{\gamma+i\Omega}{\gamma-i\Omega},\quad \mathcal K\equiv \frac{4  \gamma\,\eta^2 C_0^2}{\gamma^2+\Omega^2},\\
  \label{beta-a}
  \delta_{a+}  &=\xi\left(\epsilon_{a+} - \frac{\mathcal K\, \epsilon_{a-}}{\gamma_m-i\Omega}\right)-\\
    &\qquad         - \frac{\sqrt{\xi \mathcal K }}{\gamma_m-i\Omega}
            \left(\sqrt {2 \gamma_m} q_a + f_{s\,a}\right), \nonumber
 \end{align}
 \end{subequations}
where $\epsilon_{a\pm} = (e_{+a} \pm e_{-a})/\sqrt{2}$ represent the combinations of the input amplitude quadratures $e_{+a}$ and $e_{-a}$ of the input fields for the modes $\omega_+$ and $\omega_-$, respectively. Operators $q_a$ and $f_{s,a}$ denote the amplitude quadratures of thermal noise and the signal force acting on the mechanical oscillator, respectively.

In Eq.~\eqref{beta-a}, the back-action term is proportional to the normalized pump power $\mathcal K$, yet this contribution can be eliminated through post-processing. Specifically, by measuring both $\delta_{a+}$ and $\delta_{a-}$ simultaneously, one can completely suppress the back-action and analyze the combination
 \begin{align} \nonumber
  \delta_{meas} &= \delta_{a+} +  \frac{\mathcal K\, \delta_{a-}}{\gamma_m-i\Omega} =\\
  \label{beta-a2}
         &=\xi\epsilon_{a+} - \frac{\sqrt{\xi \mathcal K }}{\gamma_m-i\Omega}
            \left(\sqrt {2 \gamma_m} q_a + f_{s\,a}\right),
 \end{align}
which is free from back-action noise. Consequently, the precision of the force measurement is determined solely by the input fluctuations $\epsilon_{a+}$. In most cases, $\epsilon_{a+}$ describes quantum vacuum fluctuations, whose power spectral density equals unity $S_{\epsilon_{a+}} = 1$.

\begin{figure}
\includegraphics[width=0.45\textwidth]{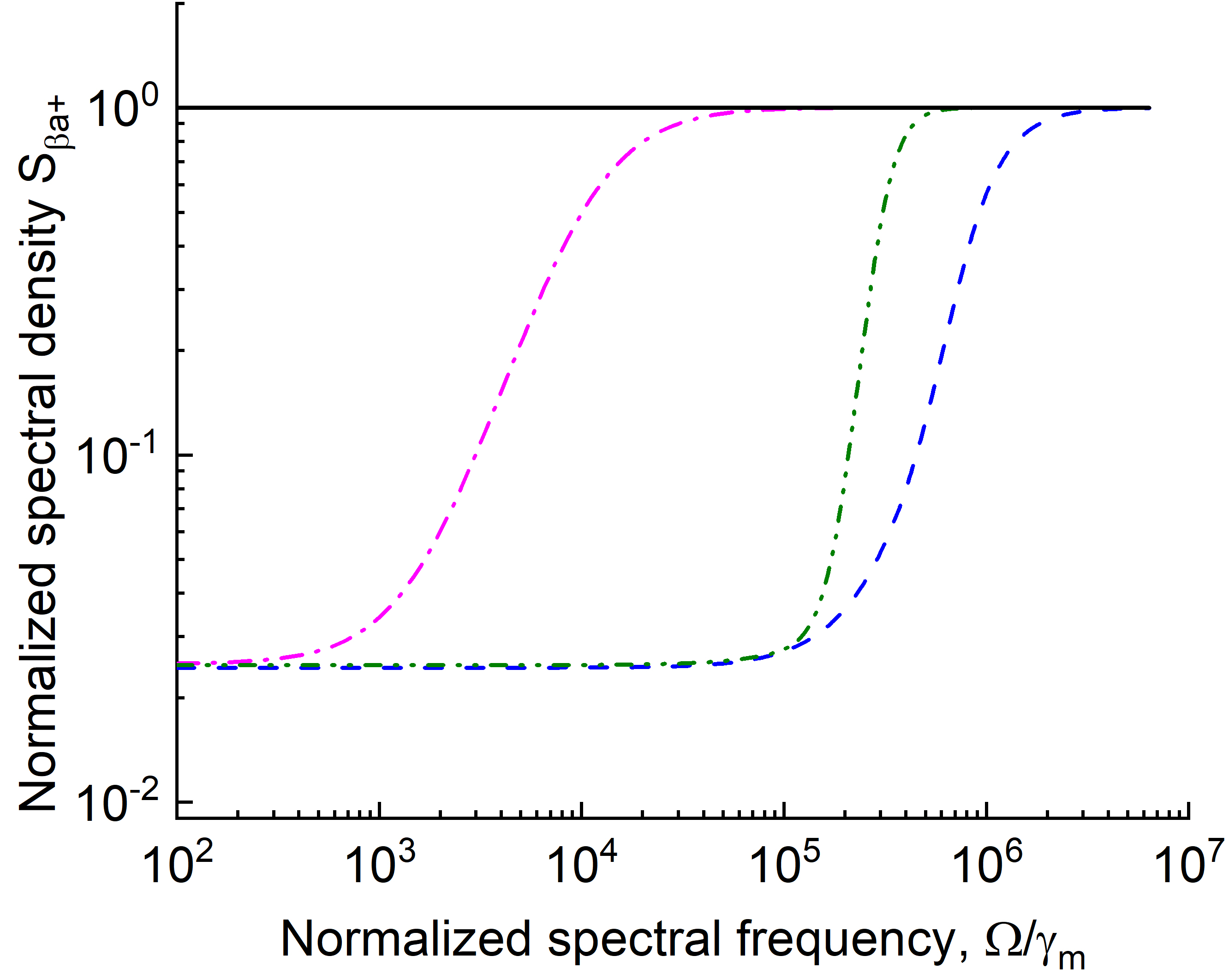}
\caption{Spectral density $S_{\beta_{a+}}$ for the parameters listed in Table~\ref{table1}.  Black line represents SQL. Dashed blue line stands for the optimal detection procedure (c.f. Fig.~\ref{fig1}) at $G/\gamma_m = 2\times 10^6,\ (G_+-G_-)/\gamma_m=10^4$.  The spectral density of the noise measured with the synodyne technique, $S_\text{syno}$, given by Eq.~\eqref{Ssyno}, is shown by the dashed-dotted magenta line at the same parameters and the dashed-dotted-dotted green line at $G/\gamma_m = 2\times 10^6,\ (G_+-G_-)/\gamma_m= \gamma_0/\gamma_m =3 \times 10^5$.}
\label{fig4}
\end{figure}

The output radiation of the configuration discussed in this paper can be directly coupled to the input of the broadband variational measurement scheme. This corresponds to
\begin{align}
\label{subs}
e_\pm = b_\pm, \quad \Rightarrow \quad \epsilon_{a+} = \beta_{a+},
\end{align}
leading to a substantially reduced spectral density:
\begin{align}
\label{SbetaSq}
S_{\epsilon_{a+}} = S_{\beta_{a+}} \ll 1.
\end{align}
In this manner, the sensitivity of the broadband variational measurement can be further enhanced.

It should be clarified that in Eq.~\eqref{EPR3} the coefficients $\theta $ and $\phi$ depend on the spectral frequency $\Omega$, which, in general, prevents a straightforward substitution as in Eq.~\eqref{subs}. However, in synodyne detection (considered previously in subsection III.C), the detected signal corresponds to a proportional combinations \eqref{EPR3} with a constant coefficient $\theta = \pi/4$ and $\phi_+ = \phi_- = 0$ (at the cost of a narrower bandwidth — see Fig.~\ref{fig4}). This demonstrates that the substitution \eqref{subs} indeed represents a two-photon squeezing process with a spectral density significantly below unity, as expressed by Eq.~\eqref{SbetaSq}.

\subsection{Stability}

\begin{figure}
\includegraphics[width=0.45\textwidth]{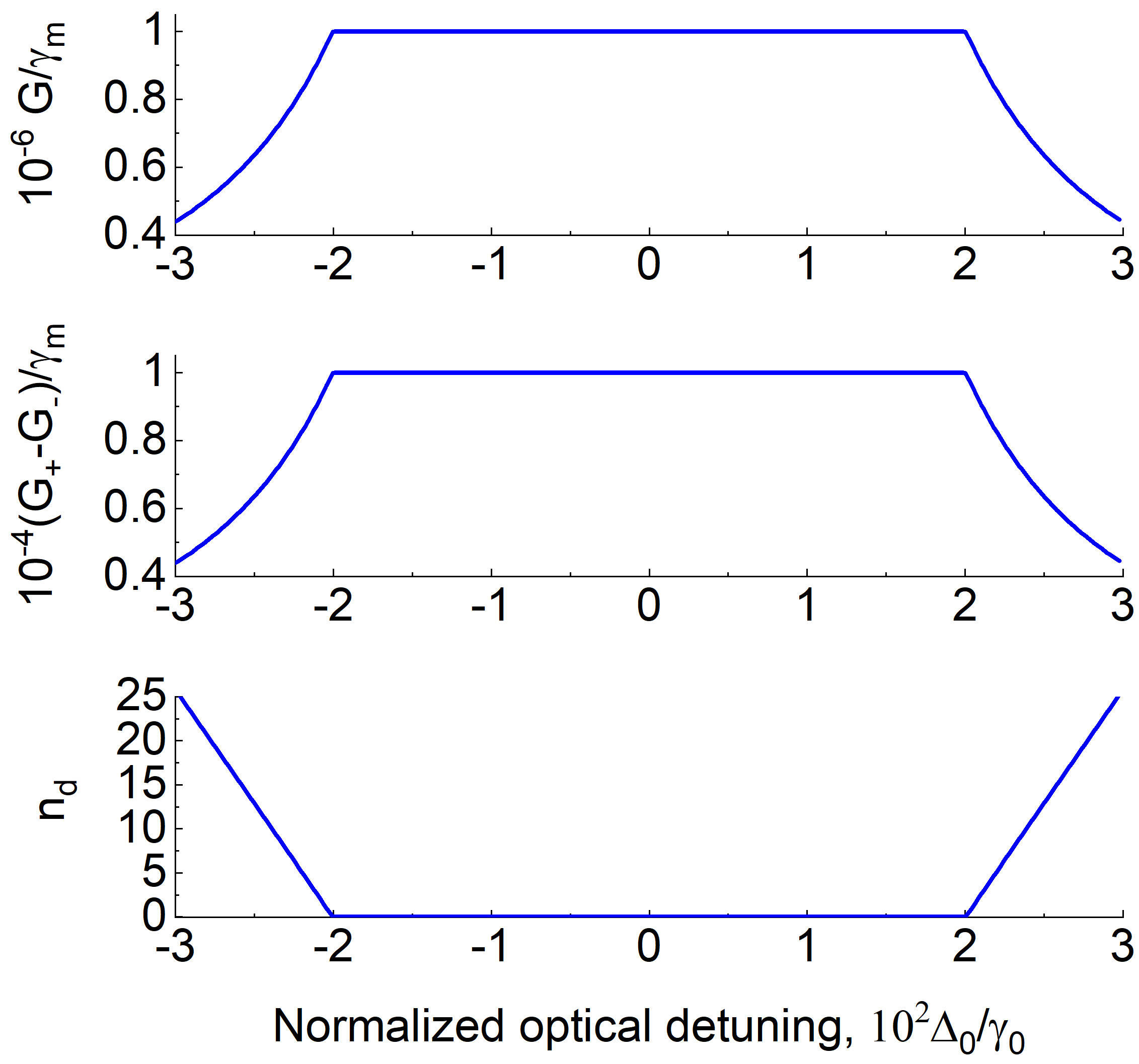}
\caption{Parameters $G$, $G_+-G_-$, and expectation value of the mechanical phonon number $n_d$ found numerically for various values of the frequency detuning $\Delta_0$  of the pump light from the corresponding optical mode. The system is stable when 
$|\Delta_0|/\gamma_0< 2\times 10^{-2}$.}
\label{fig5}
\end{figure}

The discussed opto-mechanical configuration is stable for the case of zero optical detuning even when $G\gg 1$, but $G_+>G_-$. However, introduction of nonzero detuning $\Delta_0$ of the pump from the corresponding mode results in instability. We studied the stability of the system solving numerically set (\ref{c}) for the numerical parameters listed in Table~\ref{table1} and found that the stability region exceeds the frequency bandwidth of the mechanical mode. The numerical simulation was performed by selecting the detuning value and solving the nonlinear set of equations integrating them till the steady state solution is reached. The result of simulation means that the instability can be avoided by locking the pump laser to the corresponding optical mode. The results of simulations are illustrated by Fig.~\ref{fig5}. It confirms practical viability of our proposal.

\section{Conclusion}

We investigated optomechanical entanglement between optical modes that interact via a mechanical oscillator and analyzed the feasibility of experimentally observing this quantum correlation. We demonstrate that entangled light generation requirements can be simultaneously satisfied in a realistic experimental configuration, enabling both the generation and detection of optomechanical entanglement. 

Entanglement between a mechanical oscillator and an optical field was studied in detail for the case of a single optical mode in Ref. \cite{Genes2008QEOS}. In the blue-detuned regime, such optomechanical entanglement is extremely sensitive to thermal noise: stability constraints require it to vanish once the mean thermal phonon number exceeds approximately unity. A comparable fragility with respect to thermal fluctuations has also been predicted in other multicomponent optomechanical schemes involving mechanical mediators (see Refs. \cite{riedinger2018remote, ockeloen2018stabilized}). In sharp contrast, in the configuration considered here, entanglement survives even when the mechanical mode is strongly thermally populated, including at room temperature where the mean phonon number can be many orders of magnitude larger than one.

Usually, non-linear crystals are employed to generate entanglement between two optical modes at room temperature. Compared to a conventional resonant parametric oscillator in the classical regime, the optomechanical architecture enables the generation of entangled optical modes over a significantly broader spectral bandwidth, assuming an equivalent entanglement strength. The resulting entangled light can enhance the sensitivity of optomechanical and electro-optical sensors and measurement systems. 

In systems where entanglement is generated directly between optical modes via a nonresonant optical nonlinearity, room-temperature operation is possible because thermal fluctuations of the mechanical or material degrees of freedom do not couple to the optical fields. This follows from the fluctuation–dissipation theorem: in a transparent medium the nonlinear interaction is effectively unitary and does not introduce temperature-dependent excess noise into the optical subsystem. As a result, optical entanglement can be generated without being contaminated by thermal excitations of the underlying medium.

The resonant optomechanical configuration studied here is qualitatively different. In our case, entanglement arises under a resonance condition in which the mechanical oscillator itself acts as the mediating nonlinear “medium.” One would therefore expect the large thermal occupation of the mechanical mode to severely degrade or even completely destroy the quantum correlations. Remarkably, this is not what occurs. Despite the mechanical subsystem being strongly thermally excited, robust optical entanglement is sustained, revealing a counter-intuitive regime in which resonant, thermally occupied mechanical degrees of freedom mediate rather than suppress quantum correlations between optical modes. It will be an interesting task for our future research to determine how such optomechanical entanglement can be experimentally demonstrated.

\acknowledgments
The research of AVK has been supported by Theoretical Physics and Mathematics Advancement Foundation “BASIS” (Contract No. 25-2-2-28-1). AVK and SPV would like to express their gratitude for the support provided by  the Interdisciplinary Scientific and Educational School of Moscow University ``Fundamental and Applied Space Research'' and by the TAPIR GIFT MSU Support of the California Institute of Technology. The reported here research performed by ABM was carried out at the Jet Propulsion Laboratory, California Institute of Technology, under a contract with the National Aeronautics and Space Administration (80NM0018D0004).  This document has LIGO number LIGO-P2500741.

 \bibliographystyle{ieeetr}
\bibliography{OptoMech.bib}

\end{document}